\documentclass[preprint,preprintnumbers,nofootinbib,byrevtex,prd,aps,showpacs,showkeys,groupedaddress,floatfix]{revtex4}
\usepackage{bm}
\usepackage{graphics}
\usepackage{graphicx}
\usepackage{epsfig}
\usepackage{amssymb}
\usepackage{amsmath}
\begin{document}

\large

\title{On Distribution of Zeros of Some Quazipolynoms}

\author{H.~I.~Ahmadov}
\email{hikmatahmadov@yahoo.com}
\affiliation{Department of Mathematical Physics, \\
Faculty of Applied Mathermatics and Cybernetics, \\
Baku State University Baku, Azerbaijan}

\begin{abstract}
In this paper we investigate distribution of zeros for only
quasipolynom and obtain exactly lower-bound for their modulus.
\end{abstract}

\pacs{02.30.Hq,02.70.Hm}
\keywords{spectral problem; asymptotic distribution}

\maketitle


As is known~\cite{Levin,Sadov} in connection with the
investigation of completenecess of a system of eigen and adjoint
elements of definite class of spectral problems that are not
regular by Tamarkin-Rasulov~\cite{Rasulov,Shkalikov}, there arises
the necessity of studying properties of entire analytical
functions of the form
\begin{equation}
f_k(\lambda)= e^\lambda +A_k \lambda^k
\end{equation}
($k$-is natural, and $A_k\neq 0$ - are complex constants), which
is of special interest. In Ref.~\cite{Russakovski} studied zero
sets of some classes of entire functions, gives a possibility to
describe zeros ets of certain classes of entire functions of one
and several variables in terms of growth of volume of these sets
in certain polycylinders. In Ref.~\cite{Hryniv} studied the
asymptotics of zeros for entire functions of the form
$\sin z+\int_{-1}^{1}f(t)e^{izt}dt$ with $f$ belonging to a space
$X\hookrightarrow L_{1}(-1,1)$ possessing some minimal regularity
properties. In Ref.~\cite{Ju} investigated a complete description
of zero sets for some well-known subclasses of entire functions of
exponential growth.

By taking these points into account, it may be argued that the
investigate distribution of zeros for quasipolynom.

Note that in paper~\cite{Mamedov} we have obtained the function $\Delta (\lambda )$ ($%
\Delta (\lambda )$ -is called a characteristic function) which is
an entire analytical function of a complex parameter $\lambda $
and studying a some its properties (for example, distribution of
zeros, distance between two zeros, lower estimation for modules
and ets.) is very important step in spectral theory of
differential operators.

The present paper is continuation of~\cite{Mamedov} our work and the function
of the form (1) is a special case of $\Delta (\lambda ).$

Introduce into consideration the following sets of points of the
complex surfare $C:$
$$
\Omega_{R_1R_2}(\lambda_0) = \left\{\lambda ;R_1\leq \left| \lambda
-\lambda_0\right| \leq R_2\right\}, \,\,\,\,\,\Omega_{R_1, R_2} = \Omega_{R_1, R_2}(0),
$$
$$
\Omega_R(\lambda _0) = \left\{\lambda; \left| \lambda -\lambda _0\right|
\leq R\right\}, \,\,\,\,\,\Omega_R = \Omega_{R}(0),
$$
$$
\Gamma_{kj}^S(h,R) = \left\{\lambda;\,Re\lambda +(-1)^Sk\ln \left| \lambda\right|  \right.
$$
$$
=\left.h,\,\,(-1)^jJm\lambda <0\right\} \cap \Omega_{R,\infty },
$$
$$
\Gamma_k^S(h,R) = \bigcup_{j=1}^{2}\Gamma_{kj}^S,
$$
$$
\prod\limits_{kj}^S(h,R) = \left\{ \lambda ;\,\left| Re\lambda +
(-1)^S k\ln |\lambda| \right| \leq h, (-1)^jJm\lambda <0\right\} \cap \Omega_{R,\infty },
$$
\begin{equation}
\prod\limits_k^S(h,R) = \bigcup_{j=1}^{2}\Pi _{kj}^S(h,R).
\end{equation}
$$
T_{k1}^S(h,R)=\left\{ \lambda ;\,Re\lambda +(-1)^Sk\ln |\lambda|  <-h\right\} \cap \Omega _{R,\infty },
$$
$$
T_{k2}^S(h,R)=C\backslash T_{k1}^S(h,R)\cup \Pi _k^S(h,R)\cup \overline{\Omega }_{0,R},
$$
$$
T_k^S(h,R)=\bigcup_{j=1}^{2}T_{kj}^S(h,R),
$$
$$
\Sigma _\delta ^{(i)}=\left\{ \lambda ;\,\,\left| \arg \lambda +(-1)^i\frac
\pi 2\right| <\delta \right\} ,
$$
where $0\leq R_1<R_2<\infty, \,\, R>0, \,\, h>0,\,\,i=1,2;\,\,j=1,2;\,\,
S=1,2;\,\,\delta >0.$

Now let's investigate some properties of the function $f_k(\lambda)$:
$$
\left| f_k(\lambda )\right| \geq \left| A_k\right| \left| \lambda \right|
^k\left[ 1-\left| B_k\right| e^{Re\lambda -k\ln \left| \lambda \right|
}\right] \geq
$$
$$
\left| A_k\right| \left| \lambda \right| ^k\left[ 1-\left|
B_k\right| e^{-h}\right] ,
$$
where $\left| B_k\right| =\frac 1{\left| A_k\right|}.$

In the case $\lambda \in T_{k1}^1(h,R)$ and choosing $h>\ln
2\left| B_k\right| $ we arrive at the estimation of the form
\begin{equation}
\label{3}\left| f_k(\lambda )\right| \geq \frac 12\left| A_k\right| \left|
\lambda \right| ^k,
\end{equation}
and for $\lambda \in T_{k2}^2(h,R)$ we arrive at the following
estimation of the form
\begin{equation}
\label{4}\left| f_k(\lambda )\right| \geq \frac 12\left| e^\lambda \right|
,\,\,\mbox{if} \,\,\,\,\,h>\ln 2\left|A_k\right|
\end{equation}

Thus it was proved:

Lemma 1. The function $f_k(\lambda )$ at the sufficiently large
$h>0,\,\,R>0$ in the domain $T_{k1}^1(h,R)$ and $T_{k2}^2(h,R)$
has not zeros. And the estimations (3) and (4) are true for it.
Absence of zeros of the function $f_k(\lambda )$ in the domain
$T_{k1}^1(h,R)$ and $T_{k2}^2(h,R)$ is obvions from the
estimations (3) and (4). Proceeding from the definition of the
curvilinear bands $\Pi _{kj}^S(h,R)$ the folloving is easily
proved.

Lemma 2. For any $\delta >0$ and $h>0$ we can find $R>0$, such that
$$
\prod\limits_{kj}^S(h,R)\subset \Sigma_\delta^{(1)}\cup \Sigma_\delta^{(2)}
$$
Let's prove now the following lemma.

Lemma 3. The function $f_k(\lambda )$ in the complex surfare C has
denumerable sets of zeros $\left\{ \lambda _{\nu k}\right\} $ with unique
limit point $\lambda =\infty $ which at sufficiently large $h>0,R>0$, is
situated in the domain $\Pi _k^1(h,R)\cup \overline{\Omega }_{0,R}$ . These
zeros allow the asymptotic representation:
\begin{equation}
\lambda _{\nu k}=\ln \frac{\left| A_k\right| }{\left[ 2\pi \left|
\nu \right| \right] ^k}+i\left( 2\pi \nu +\pi +\frac{\pi k}2+
\arg A_k\right)+0\left( \frac{\ln \left| \nu \right| }\nu \right)
\label{5}
\end{equation}
Proof. The assertion of the first part of the lemma follows from the general theory of
Picard~\cite{Privalov} and from the lemma 1. Prove the second part of the lemma.
\newpage
$$
f_k(\lambda )=0, \,\,\,\,\,\,\,\,e^\lambda +A_k\lambda ^k=0,
$$
$$
e^\lambda \cdot \lambda ^{-k}=-A_k,\,\,\,\,\,\,\,\,e^\lambda \cdot e^{-k\ln \lambda }=-A_k,
$$
$$
\lambda -k\ln \left| \lambda \right| =\ln \left| A_k\right| +i\left( \arg
(-A_k)+2\pi \nu \right) .
$$
Make the substitution $\lambda -2\pi \nu i=\xi _\nu ,$
then
$$
\xi _\nu =\ln \left| A_k\right| +i\arg (-A_k)+k\ln \lambda =
$$
$$
\ln \left| A_k\right| +i\left( \arg A_k+\pi \right) +k\ln \left( 2\pi \nu
i+\xi _\nu \right) =
$$
$$
\ln \left| A_k\right| +i\left( \arg A_k+\pi \right) +
k\ln \left[ 2\pi \nu i\left( 1+\frac{\xi _\nu }{2\pi \nu i}\right) \right] =
$$
$$
\ln \left| A_k\right| +i\left( \arg A_k+\pi \right) +k\ln (2\pi \nu i)+
k\ln\left( 1+\frac{\xi _\nu }{2\pi \nu i}\right).
$$
Since
$$
\lim_{\nu \to \infty }\ln \left( 1+\frac{\xi _\nu }{2\pi \nu i}\right) =0,\,
\ln \,\,\left( 1+\frac{\xi _\nu }{2\pi \nu i}\right) =
$$
$$
0\left( \frac{\xi _\nu }{2\pi \nu i}\right) =0\left( \frac{\ln \left| \nu
\right| }\nu \right) .
$$
$$
\xi _\nu =\ln \left| A_k\right| +i\left( \arg A_k+\pi \right) +k\ln (2\pi
\nu i) + O\left( \frac{\ln \left| \nu \right| }\nu \right) ,
$$
then we find
$$
\lambda _\nu =\ln \frac{\left| A_k\right| }{\left[ 2\pi \left| \nu \right|
\right]^{-k}}+i\left( 2\pi \nu +\pi +\frac{k\pi}{2} + \arg A_k\right) +O\left(\frac{\ln \left| \nu \right| }\nu \right) .
$$
From (5) we see that at large $\left| \nu \right| $ we have
\begin{equation}
\label{6}\left| \lambda_{\nu +1}-\lambda_\nu \right| =2\pi +0(1)
\end{equation}
Consequently there exists $\delta >0$ such that the circles of $\Omega
_\delta (\lambda _\nu )$ are mutually exclusive and at the sufficiently large
$h>0, R>0$ wholly lie in the domain $\prod\limits_k^1(h,R)\cup \overline{\Omega}_{0,R}.$
From the asymptotic formulae (5) and (6) we see that straight lines
$$
l_{\nu ,k}=\left\{ \lambda :\,\,Jm\lambda =Jm\lambda _\nu - \left( \pi +\frac{\pi k}{2}+\arg A_k\right) \right\}
$$
are perpendicular to the imaginary axis $Re\lambda =0$ at all
possible different, sufficiently large (by the modulus) values of
$\nu $, are different. This lines divide the domains
$\prod\limits_k^1(h,R)$ into the curvilinear quadrangles $D_{\nu k}=D_{\nu k}(h,R)$
with lateral boundaries on the lines $\gamma_k(-h,R),\,\gamma _k(h,R)$
and lies on the straight lines $l_{\nu-1,k}>l_{\nu ,k}$.
The length of the diagonal of a quadrangle denote by
$$
d_\nu =\sup_{\lambda ,\mu \in D_{\nu ,k}} \left| \lambda -\mu
\right| .
$$
Introduce the following notation
$$\prod\limits_k^1(h,R,\delta )=
\prod_k^1(h,R) \setminus \bigcup\limits_{\nu}\Omega _{0,\delta }(\lambda_{\nu}),$$
where the sigh of unification is propagated on all $\nu $ such that
$$
\lambda _\nu \in \prod\limits_k^1(h,R),
$$
$$
D_{k\nu }^\delta =D_{k\nu }\setminus \Omega _{0,\delta }(\lambda _\nu ).
$$
The following lemma is true.

Lemma 4. There exists the constant $\delta >0$ , such that at
$\lambda \in \prod\limits_k^1(h,R,\delta )$ it holds the inequality
\begin{equation}
\label{7}\left| f_k(\lambda )\right| \geq C_\delta \left| \lambda \right|^k.
\end{equation}
%

\end{document}